\documentclass[a4paper,11pt]{article}

\usepackage{jheppub} 

\usepackage[T1]{fontenc} 
\usepackage{lmodern}

\usepackage{tikz}
\usetikzlibrary{positioning,decorations.pathmorphing}

\let\oldFootnote\footnote
\newcommand\nextToken\relax

\renewcommand\footnote[1]{%
    \oldFootnote{#1}\futurelet\nextToken\isFootnote}

\newcommand\isFootnote{%
    \ifx\footnote\nextToken\textsuperscript{,}\fi}

\usepackage{enumerate}

\usetikzlibrary{decorations.markings}

\def\id{{1 \kern-.28em {\rm l}}}

\def\K3{{\bf K3}}
\def\journal#1&#2(#3){\unskip, \sl #1\ \bf #2 \rm(19#3) }
\def\andjournal#1&#2(#3){\sl #1~\bf #2 \rm (19#3) }

\def\bar{\overline}

\def\ie{{\it i.e.}}
\def\eg{{\it e.g.}}

\def\etc{{\it etc}}

\def\tilde{\widetilde}

\def\frac#1#2{{#1\over#2}}

\def\inbar{\,\vrule height1.5ex width.4pt depth0pt}
\def\IC{\relax\hbox{$\inbar\kern-.3em{\rm C}$}}
\def\IR{\relax{\rm I\kern-.18em R}}
\def\IP{\relax{\rm I\kern-.18em P}}

\catcode`\@=11
\def\slash#1{\mathord{\mathpalette\c@ncel{#1}}}
\def\underrel#1\over#2{\mathrel{\mathop{\kern\z@#1}\limits_{#2}}}

\catcode`\@=12

\def\tr{{\rm tr}}

\def\exp{{\rm exp}}

\def\ie{{\it i.e.}}
\def\eg{{\it e.g.}}

\preprint{TIFR/TH/20-28}
\title{ $T\bar{T}$ and $J\bar{T}$ Deformations in Quantum Mechanics}

\author{Soumangsu Chakraborty,  Amiya Mishra}
\emailAdd{soumangsuchakraborty@gmail.com}
\emailAdd{mishramiya@gmail.com}

\affiliation{Department of Theoretical Physics,\\Tata Institute of Fundamental Research,\\Homi Bhabha Road, Mumbai 400005, India}

\abstract{In this paper, we continue the study of $T\bar{T}$ deformation in $d=1$ quantum mechanical systems and propose possible analogues of $J\bar{T}$ deformation and deformation by a general linear combination of $T\bar{T}$ and $J\bar{T}$ in quantum mechanics. We construct  flow equations for the partition functions of the deformed theory, the solutions to which yields the deformed partition functions as integral of the undeformed partition function weighted by some kernels. The kernel formula turns out to be very useful in studying the deformed two-point functions and analyzing the thermodynamics of the deformed theory. Finally, we show that a non-perturbative UV completion of the deformed theory is given by minimally coupling the undeformed theory to worldline gravity and $U(1)$ gauge theory. }

\begin{document}
\maketitle
\flushbottom

\section{Introduction}

Over the past few years, there has been a growing interest in understanding irrelevant deformations of a $CFT_2$, the most well-studied among them are the $T\bar{T}$ \cite{Smirnov:2016lqw,Cavaglia:2016oda}, $J\bar{T}$ \cite{Guica:2017lia,Chakraborty:2018vja} and deformation by their general linear combination \cite{Chakraborty:2019mdf,LeFloch:2019rut}. Here, $T$ and $\bar{T}$ are the holomorphic and anti-holomorphic components of the stress tensor and $J$ is the holomorphic component of a global $U(1)$ current. Although such deformations are irrelevant, involving flowing up the renormalization group (RG) trajectory, the deformations turn out to be integrable \cite{Smirnov:2016lqw} and one can compute the exact spectrum of the deformed theory in terms of the undeformed spectrum and the couplings. The resulting theories in the UV are non-local in the sense that their short distance behavior is not governed by  fixed points.

It is well-known that for one sign of the $T\bar{T}$ coupling, the deformed spectrum is non-unitary. Spectrum of states with energies above some critical value in the undeformed theory, upon deformation becomes complex. For this particular sign of the coupling, there has been a proposal for the holographic dual in the form of $AdS_3$ with a hard radial cutoff \cite{McGough:2016lol}.\footnote{For other holographic proposals and their generalizations, see \cite{Giveon:2017nie,Chakraborty:2018vja,Chakraborty:2019mdf,Chakraborty:2020swe,Guica:2019nzm}.} In an attempt to visualize holography with a hard Dirichlet wall in higher dimensions, the authors of  \cite{Bonelli:2018kik,Taylor:2018xcy,Hartman:2018tkw,Belin:2020oib} constructed a dual effective field theory which in the deep IR appears to be a large $N$ CFT  deformed by an irrelevant operator which is bilinear in the components of the stress tensor. Such a deformation is often regarded as an analogue of $T\bar{T}$ deformation in higher dimensions.\footnote{In this paper, by abuse of language, we will keep referring to these deformations in $d>2$ and in $d=1$ as the $T\bar{T}$ deformation, similarly for $J\bar{T}$ and their general linear combinations.} The $T\bar{T}$ operator being a composite operator needs to be defined by point splitting. In $d=2$ it has been proved that such an operator is indeed well-defined at the coincident limit \cite{Zamolodchikov:2004ce}. In higher dimensions, however, the well-definedness at the coincident limit is yet to be proven.

Motivated from the holographic analysis with a hard UV wall in dimensions $d\geq 2$, the authors of \cite{Gross:2019ach,Gross:2019uxi} defined an analogue of $T\bar{T}$ deformation in $d=1$ from a thorough analysis of JT gravity in $AdS_2$ with a hard radial cutoff (see also \cite{Iliesiu:2020zld}). In the absence of any spatial directions in $d=1$ quantum mechanics, the $T\bar{T}$ operator defined by the right hand side of the flow equation \eqref{ttbarflow},  is a composite operator and is well-defined by construction. Such an operator is irrelevant and deformation of a quantum mechanical system in the IR by this operator involves flowing up the RG trajectory much like its higher dimensional cousins. The thermodynamic stability of $T\bar{T}$ deformation in $d=1$ has been studied in \cite{Barbon:2020amo}.

However, RG flows in one dimension enjoy many properties that are absent in higher dimensions. In $d=1$, there exists only a handful of irrelevant operators in contrast to a gigantic set of relevant operators. This behavior is quite opposite to what happens in $d\geq 2$. As a result, the space of theories in the IR is very rich compared to the landscape in the UV which is highly universal. In some sense, the ambiguities in flowing up the RG triggered by some irrelevant deformation in $d\geq 2$ doesn't exist in $d=1$.  This itself is a good enough motivation to study irrelevant deformations like $T\bar{T}$ and $J\bar{T}$ deformation in $d=1$ quantum mechanical systems.

In this paper, we continue the study of $T\bar{T}$ deformation in $d=1$ initiated in \cite{Gross:2019ach} and introduce an analogue of $J\bar{T}$ deformation in quantum mechanics and eventually study deformation by general linear combination of $T\bar{T}$ and $J\bar{T}$. We have shown that the deformed spectrum has the same square root brunch cut singularity as in the case of their higher dimensional cousins. In the case of pure $J\bar{T}$ deformation, we have shown that the deformed spectrum is non-unitary for either sign of the coupling as in $d=2$ and upon switching on the $T\bar{T}$ deformation we see that there is a region in the parameter space (\ie\ the space of $T\bar{T}$ and $J\bar{T}$ coupling) where the theory is unitary. We introduce a kernel formula such that the integral of the undeformed partition function weighted by some kernel gives rise to the deformed partition function. As an application of the kernel formula, we compute the deformed two-point functions of operators\footnote{We have considered only those operators that do not develop explicit dependence on the coupling upon deformation.} of the deformed theory. Then, we analyze the thermodynamics of a certain class of theories (systems with linear specific heat) deformed by a general linear combination of $T\bar{T}$ and $J\bar{T}$ and  show that there is a region in the parameter space, where the theory is unitary and exhibits Hagedorn behavior, similar to the two dimensional case \cite{Chakraborty:2020xyz}. Unlike $d=2$, the Hagedorn behavior at high energies is typical to systems with linear specific heat (\eg\ Schwarzian theories \cite{Stanford:2017thb}). Systems with non-linear specific heat are not likely to exhibit Hagedorn growth. 

It has been shown in \cite{Gross:2019ach}, that the non-perturbative UV completion of $T\bar{T}$ deformed quantum mechanics is obtained by coupling the theory to one-dimensional worldline gravity. This can be realized as the one-dimensional version of the non-perturbative UV completion of $T\bar{T}$ deformed quantum field theory in two dimensions where the deformed theory is obtained by minimally coupling the undeformed theory with JT gravity in flat space \cite{Dubovsky:2018bmo}. In the same spirit, we show a possible non-perturbative UV completion of quantum mechanics deformed by a general linear combination of $T\bar{T}$ and $J\bar{T}$ is obtained by coupling the undeformed theory to worldline gravity and   $U(1)$ gauge theory.

\subsubsection*{Plan of the paper}

The organization of this paper is as follows. In section \ref{sec2}, we give a brief review of $T\bar{T}$, $J\bar{T}$ and $T\bar{T}+J\bar{T}$ deformation and their deformed spectrum in two dimensions. In section \ref{sec3}, we define an analogue of $T\bar{T}$, $J\bar{T}$ and $T\bar{T}+J\bar{T}$ deformation in quantum mechanics through their respective flow equations and derive the kernel formulas to compute the deformed partition functions.  In section \ref{sec4}, we compute the deformed two-point function and analyze the thermodynamics of the deformed theory. We also comment on the thermodynamic stability of the deformed theory. In section \ref{sec5}, we discuss a possible non-perturbative UV completion of the deformed theory. Finally, in section \ref{sec6}, we conclude our findings and discuss various outlooks and potential future directions.

\section{Brief review of $T\bar{T}$ and related deformations in $d=2$}\label{sec2}

In this section, we will present a brief review of what it means by deformation of a $d=2$ QFT by the operators $T\bar{T}$, $J\bar{T}$ and a linear combination of $T\bar{T}$, $J\bar{T}$. Here, we will be brief and  sketch only the expression of the deformed spectrum without going into much details. 

\subsection{$T\bar{T}$ deformation}
Let us consider a $d=2$ Lorentz invariant $QFT$ on a cylinder $\mathbb{R}\times S^1$ of radius $R$ with  a stress tensor $T_{\mu\nu}$ which is conserved. Conservation  implies
\begin{eqnarray}
\partial_\mu T^{\mu\nu}=0.
\end{eqnarray}
 Then, the $T\bar{T}$ operator of such a system is defined as
\begin{eqnarray}
T\bar{T}=-{\rm{det}} (T_{\mu\nu}).
\end{eqnarray}
Let the Lagrangian density of such a theory (before deformation) be denoted by $\mathcal{L}_0$.
Note that the $T\bar{T}$ operator is a composite operator and must be defined by point splitting. That the coincident limit is finite up to total derivative terms has been proved in \cite{Zamolodchikov:2004ce}. 

Next, let us consider an RG flow in the space of theories parametrized by the affine parameter $\lambda$ such that at each point on the trajectory in the space of theories, the flow is triggered by the operator $T\bar{T}(\lambda)$ at that point on the trajectory. Setting $\lambda=0$ gives back the seed theory one started with. 

Let the Lagrangian of the deformed theory at an arbitrary point on the RG trajectory be given by $\mathcal{L}_\lambda$. Then the deformation of a $d=2$ Lorentz invariant $QFT$ by the operator $T\bar{T}$ is given by the flow equation
\begin{eqnarray}
\frac{\partial \mathcal{L}_\lambda}{\partial \lambda}=-\frac{1}{2}T\bar{T}(\lambda)~.
\end{eqnarray}
Although this deformation is irrelevant, it turns out to be solvable. The deformed spectrum, $E(\lambda; E_0,P)$, can be expressed as a function of the spectrum $E_0$  and momentum $P$ of the undeformed theory as
\begin{equation}\label{TTbarspec}
E(\lambda; E_0,P)=-\frac{R}{\lambda}\left(1-\sqrt{1+\frac{2\lambda E_0}{R}+\frac{\lambda^2P^2}{R^2}}\right).
\end{equation}
Note that for $\lambda>0$, the deformed spectrum is unitary as opposed to for $\lambda<0$, in which case the spectrum becomes complex above some threshold.

\subsection{$J\bar{T}$ deformation}

To define a $J\bar{T}$ deformation of a CFT, let us start with a $d=2$ CFT on a cylinder of radius $R$ that contains a conserved left moving $U(1)$ current $J$. Then the flow equation of  $J\bar{T}$ deformation is given by
\begin{eqnarray}
\frac{\partial \mathcal{L}_\alpha}{\partial \alpha}=2J(\alpha)\bar{T}(\alpha),
\end{eqnarray}
where $\alpha$ is the coupling,  $J(\alpha)$ is the deformed left-moving $U(1)$ current and $\bar{T}(\alpha)$ is the deformed right-moving component of the stress tensor. Similar to the $T\bar{T}$ deformation, the $J\bar{T}$ deformation is irrelevant, but unlike the previous case, $J\bar{T}$ deformation breaks Lorentz invariance.

It has been argued in \cite{Chakraborty:2018vja} that all along the RG flow,
\begin{enumerate}[(i)]
 \item the left-moving component of the stress tensor, $T(\alpha)$, remains holomorphic \ie\   $\bar{\partial}T(\alpha)=0$, 
 \item the left moving component of the global $U(1)$ current remains holomorphic \ie\ $\bar{\partial}J(\alpha)=0$,
 \item the quantity $T(\alpha)-\frac{1}{2}J(\alpha)^2$ is independent of the coupling $\alpha$.
 \end{enumerate}
  From the above three conditions one can construct the deformed spectrum that takes the following form \cite{Guica:2017lia,Chakraborty:2018vja}:
\begin{equation}\label{jtspect}
E(\alpha;E_0,P,Q)=P+\frac{2R}{\alpha^2}\left(\left(1-\alpha\frac{ Q}{R}\right)-\sqrt{\left(1-\alpha \frac{Q}{R}\right)^2-\frac{\alpha^2}{R}\left(E_0-P\right)}\right),
\end{equation}
where, $Q$ is the undeformed left-moving $U(1)$ charge. Note that in this case the spectrum above some threshold, becomes complex for both signs of the coupling. 

\subsection{Deformation by a general linear combination of $T\bar{T}$, $J\bar{T}$ and $T\bar{J}$}

Next, one can consider a deformation by a general linear combination of $T\bar{T}$, $J\bar{T}$ and $T\bar{J}$.
The flow equations are given by
\begin{eqnarray}
\frac{\partial\mathcal{L}_{\lambda,\alpha_\pm}}{\partial{\lambda}}&=&-\frac{1}{2}T\bar{T},\\
\frac{\partial\mathcal{L}_{\lambda,\alpha_\pm}}{\partial{\alpha_+}}&=&2J\bar{T},\\
\frac{\partial\mathcal{L}_{\lambda,\alpha_\pm}}{\partial{\alpha_-}}&=&2T\bar{J},
\end{eqnarray}
where, $\alpha_{\pm}$ are respectively the $J\bar{T}$ and $T\bar{J}$ coupling.
Defining the theory is a bit involved, so we will refer to the references \cite{LeFloch:2019rut,Chakraborty:2019mdf} for more details. The deformed spectrum takes the following form
\begin{eqnarray}
E(\lambda,\alpha_{\pm},E_0,P,Q_{L,R})=P-\frac{1}{2AR}\left(-B-\sqrt{B^2+4AC}\right),\label{TTpJTspect}
\end{eqnarray}
with,
\begin{eqnarray}
A&=&\frac{1}{4R^2}\left(2\lambda-(\alpha_++\alpha_-)^2\right), \nonumber\\
B&=& -1+\frac{1}{R}(\alpha_+Q_L+\alpha_-Q_R)+\frac{P}{R}\alpha_-(\alpha_++\alpha_-)-\frac{\lambda P}{R},\label{ABC}\\
C&=& (E_0-P)R+2\alpha_-Q_RP+\alpha_-^2P^2,\nonumber
\end{eqnarray}
where, $Q_{L,R}$ are the left and right-moving $U(1)$ charges. For $A\geq0$ the deformed spectrum is unitary, else there is a threshold above which the spectrum becomes complex.

In the following section, we are going to draw motivations from the expressions of the deformed spectrum in $d=2$ to define an analogue of $T\bar{T}$ and $J\bar{T}$   deformations in one-dimensional quantum mechanical system.

\section{$T\bar{T}$ and related deformation in quantum mechanics}\label{sec3}

The authors of \cite{Gross:2019ach,Gross:2019uxi} discussed an infinite class of one parameter family of solvable deformations of quantum mechanical systems where the undeformed Hamiltonian, $H_0$, gets mapped to some function of itself \ie\ $H=f(\lambda,H_0)$ where, $H$ denotes the deformed Hamiltonian and $H(\lambda=0)=H_0$. Since $[H_0,H]=0$, the energy eigenstates remain undeformed under the deformation. The eigenvalues of the deformed Hamiltonian, on the other hand, change as $E=f(\lambda, E_0)$, where $E_0$ is the spectrum of the undeformed system. Such a deformation is defined by a flow equation 
\begin{equation}
\frac{\partial H}{\partial \lambda}=F(\lambda,H),\label{diff}
\end{equation}
where, $F(\lambda,H)$ is the deforming operator and $\lambda$ is the deformation parameter. Note that \eqref{diff} is a first order differential equation implying that its solution is unique given an initial condition.  

These solvable deformations have an obvious generalization to systems that have additional commuting global symmetries. In that case the deformed Hamiltonian takes the form $H=f(\lambda, H_0,Q)$ where $Q$ is the charge associated with the additional global symmetry.\footnote{Here we are assuming that $Q$ remains undeformed upon deformation.}  An example of such a deformation would be the analogue of $J\bar{T}$ deformation in quantum mechanics that we will discuss in details in the subsection \ref{JTbarqm}.  

These solvable deformations have a further generalization to two or more parameter family of flow equations namely
\begin{eqnarray}
\frac{\partial H}{\partial\lambda_i}=F_i(\{\lambda_i\},H), \ \text{ with } \ i=1,2,\cdots.
\end{eqnarray}
An example of such a deformation would be deformation by a general linear combination of $T\bar{T}$ and $J\bar{T}$ in quantum mechanical systems with a global $U(1)$ symmetry, which we are going to introduce in subsection \ref{TTpJT}.

\subsection{$T\bar{T}$ deformation in quantum mechanics}\label{TTbarqm}

The motivation for studying $T\bar{T}$ deformation in quantum mechanics is two-fold. Firstly, to understand $JT$ gravity with a finite radial cutoff in the bulk $AdS_2$ \cite{Gross:2019ach,Gross:2019uxi,Iliesiu:2020zld} and secondly to study a solvable deformation of quantum mechanics that would give rise to a deformed spectrum with a square root branch cut singularity like the one that appears in two dimensions \eqref{TTbarspec}. The $T\bar{T}$ flow equation as defined in  \cite{Gross:2019ach,Gross:2019uxi} takes the following form:
\begin{eqnarray}
\frac{\partial S_E}{\partial \lambda}=-\int d\tau ~ \frac{T^2}{2+2\lambda T},\label{ttbarflow}
\end{eqnarray}
where, $S_E$ is the euclidean action of the deformed theory, $\tau$ is the Euclidean time, and $T=T^{\tau}_{\tau}=H$
is the stress tensor of the system. This leads to the differential equation for the deformed spectrum
\begin{equation}\label{flowttbar}
2\frac{dE}{d\lambda} +2\lambda E\frac{dE}{d\lambda} + E^2 =0,
\end{equation}
which solves as
\begin{equation}\label{sect1tt}
E(\lambda)=- \frac{1}{\lambda}\left(1-\sqrt{1+2\lambda E_0} \right).
\end{equation}
The spectrum agrees precisely with the energy of a two dimensional black hole in $AdS_2$ with a finite radial cutoff \cite{Gross:2019ach} for $\lambda<0$. Note that, the limit $\lambda\to0$ leads to the undeformed spectrum. As in the case of two dimensions, the deformed spectrum is unitary for $\lambda>0$ whereas for $\lambda<0$, there is a threshold above which the spectrum becomes complex.   One can thus read off the  deformed Hamiltonian operator as 
\begin{eqnarray}
H(\lambda)=- \frac{1}{\lambda}\left(1-\sqrt{1+2\lambda H_0} \right).
\end{eqnarray}

\subsubsection{Thermal partition function and its flow equation}
In this subsubsection, we are going to present an alternative way to realize the flow equations in terms of the thermal partition function eventually leading to a kernel formula that generates the deformed thermal partition function. This method turns out to be more useful in analyzing the deformed thermal partition function in the case of  $J\bar{T}$ deformation and eventually to a class of  more general deformations.

The thermal partition function of the deformed theory is given by the celebrated trace formula
\begin{eqnarray}
Z(\lambda,\beta)=\tr\left(e^{-\beta H(\lambda)}\right)=\sum e^{-\beta E(\lambda)},
\end{eqnarray}
where, $\beta$ is the inverse temperature. Note that $\beta$ can also be realized as the periodicity of the Euclidean time $\tau$ namely $\tau\sim\tau+\beta$.

Equation \eqref{flowttbar} implies 
\begin{equation}
\tr\left[\left(2\frac{dH}{d\lambda} +2\lambda H\frac{dH}{d\lambda} + H^2\right)e^{-\beta H}\right]=0. \label{par1}
\end{equation}
The expectations of the individual terms on the l.h.s. of \eqref{par1} take the following forms:
\begin{equation}
Z\left\langle \frac{dH}{d\lambda}\right \rangle  =-\frac{1}{\beta}  \frac{\partial Z}{\partial \lambda} , \ \ \ \ \
Z\left\langle H \frac{dH}{d\lambda} \right\rangle =\frac{1}{\beta}  \bigg(\frac{\partial }{\partial \beta} -  \frac{1}{\beta} \bigg) \frac{\partial Z}{\partial \lambda}, \ \ \ \ \ 
Z\left\langle H^2 \right\rangle = \frac{\partial Z^2}{\partial \beta^2}.
\label{expZtt}
\end{equation}
Substituting \eqref{expZtt} in \eqref{par1} gives the flow equation in terms of the thermal partition function:
\begin{equation}
-\frac{2}{\beta}  \frac{\partial Z}{\partial \lambda} +\frac{2\lambda}{\beta} \bigg(\frac{\partial }{\partial \beta} -  \frac{1}{\beta} \bigg) \frac{\partial Z}{\partial \lambda}+ \frac{\partial Z^2}{\partial \beta^2} =0, \label{flowpartt}
\end{equation}
which matches exactly with the flow equation constructed from the bulk JT gravity in cutoff $AdS_2$ with Dirichlet boundary condition on the cutoff surface \cite{Iliesiu:2020zld}. This can also be visualized as the defining equation for $T\bar{T}$ deformation in quantum mechanics. Equation \eqref{flowpartt} solves as
\begin{equation}\label{kerfortt}
Z(\lambda,\beta)= \int_{\mathcal{C}_\beta} d\beta' \ K(\lambda,\beta,\beta') ~ Z_0(\beta'),
\end{equation}
where $Z_0(\beta') $ is the partition function of the undeformed theory at temperature $1/\beta'$, the contour $\mathcal{C}_\beta$ runs from $0$ to $\infty$ along the real $\beta$ axis  and the kernel $K(\lambda,\beta,\beta')$ is given by \footnote{If the quantum mechanical system has a finite chemical potential then kernel \eqref{kertt} can be trivially generalized by multiplying it with $\delta(\mu'-\mu)$ (where $\mu'$ and $\mu$ are respectively the chemical potentials of the undeformed and deformed systems.) and integrating over $\mu'$ in \eqref{kerfortt} from $-\infty$ to $\infty$.}
\begin{equation}\label{kertt}
K(\lambda,\beta,\beta') =\frac{\beta}{\beta'^{3/2}\sqrt{2\pi\lambda}}~\exp\bigg(-\frac{(\beta-\beta')^2}{2\beta'\lambda} \bigg).
\end{equation}
The kernel agrees with the one presented in \cite{Gross:2019uxi}. It is obvious that the kernel \eqref{kertt} also satisfies the differential equation \eqref{flowpartt}. From \eqref{kerfortt} one can read off the spectrum of the deformed theory which turns out to be \eqref{sect1tt}.

Note that the kernel formula \eqref{kerfortt} with \eqref{kertt} is strictly well-defined for $\lambda>0$. For $\lambda<0$ the integral in \eqref{kerfortt} is divergent. This is related to the fact that for $\lambda<0$, the spectrum is non-unitary. We will hence take the attitude of defining the integral for $\lambda>0$ and  analytically continue the final result for $\lambda<0$.\footnote{Alternatively, for $\lambda<0$ one can choose a separate contour such that the $\beta'$ integral is convergent.}

\subsection{$J\bar{T}$ deformation in quantum mechanics}\label{JTbarqm}

An obvious question that one may raise at this point is:  what would be an analogue of $J\bar{T}$ deformation in quantum mechanics?  In the absence of any nice holographic interpretation, we would like to draw motivation from the spectrum \eqref{jtspect} of a $CFT_2$ deformed by the $J\bar{T}$ operator. The deformed spectrum in two dimensions has the properties that it has a square root branch cut much like the case of $T\bar{T}$ and  that for both signs of the $J\bar{T}$ coupling, the spectrum is non-unitary. Inspired from  
 \eqref{jtspect}, let us define the deformed spectrum of a one-dimensional quantum mechanical system with a global $U(1)$ charge $Q$ (which in quantum mechanics is also the conserved current $J$) as follows:\footnote{Note that in quantum mechanics, there are no space directions. Thus in defining the deformed spectrum in one dimension, one sets the momentum $P=0$. }
 \begin{equation}\label{spectjt}
E(\alpha)=\frac{2}{\alpha^2} \left(1- \frac{\alpha Q}{\sqrt{\ell}}-\sqrt{\left(1-\frac{\alpha Q}{\sqrt{\ell}} \right)^2-\alpha^2 E_0}~ \right),
\end{equation}
where $\alpha$ is the coupling of the deforming operator, and $\ell$ is some length scale associated with the field space of undeformed theory (\eg\ for a free compact scalar $y\sim y+2\pi \sqrt{\ell}$, $\sqrt{\ell}$ is the radius of the compact boson in the configuration space.).\footnote{In the canonical formalism, the charge $Q$ is dimensionless in all dimensions. The coupling $\alpha$ has length dimension $1/2$. Thus to make the quantity $\alpha /\sqrt{\ell}$ dimensionless, $\ell$ must have dimension of length. }

The flow equation that would give rise to such a deformed spectrum is given by 
\begin{equation}
\frac{\partial S_E}{\partial\alpha}=\int d\tau~\frac{\alpha T^2+\frac{2JT}{\sqrt{\ell}}}{2\left(1-\frac{\alpha J}{\sqrt{\ell}}\right)-\alpha^2T}.
\end{equation}
This leads to the following differential equation of the deformed spectrum
\begin{equation}\label{flowjtbar}
2\frac{dE}{d\alpha} - \frac{2\alpha Q}{\sqrt{\ell}} \frac{dE}{d\alpha}  -\alpha^2 E \frac{dE}{d\alpha} -\frac{2 Q}{\sqrt{\ell}}E-\alpha E^2=0,
\end{equation}
which solves as \eqref{spectjt}.\footnote{Note that, $Q=0$ and $\alpha^2\to-2\lambda$ give rise to the flow equation and deformed spectrum for the $T\bar{T}$ case.}\footnote{Recall that we assumed $Q$ remains undeformed all along the flow.}

\subsubsection{Thermal partition function and its flow equation}
Next, we are going to define a flow equation in terms of the thermal partition function. Here we follow the same line of arguments as used in the case of $T\bar{T}$ deformation. The thermal partition function of the deformed system coupled to a finite chemical potential $\mu$ is given by
\begin{eqnarray}
Z(\alpha,\beta,\mu)= \tr\left(e^{-\beta H+\mu Q/\sqrt{\ell}}\right).
\end{eqnarray} 
The flow equation of the deformed spectrum \eqref{flowjtbar} implies
\begin{equation}
\tr\left[\left(2\frac{dH}{d\alpha} - \frac{2\alpha J}{\sqrt{\ell}} \frac{dH}{d\alpha}  -\alpha^2 H \frac{dH}{d\alpha} -\frac{2 J}{\sqrt{\ell}}H-\alpha H^2\right)e^{-\beta H+\mu Q/\sqrt{\ell}}\right]=0.\label{part2}
\end{equation}
The expectation of the individual terms on the l.h.s. of \eqref{part2} take the following forms:
\begin{equation}\label{exphj}
\begin{split}
Z\langle \frac{dH}{d\alpha} \rangle  =-\frac{1}{\beta}  \frac{\partial Z}{\partial \alpha} ,\ \  & \ \  Z\langle J \frac{dH}{d\alpha} \rangle  =-\frac{\sqrt{\ell}}{\beta}  \frac{\partial^2 Z}{\partial \mu \partial \alpha}  , \\
Z\langle H \frac{dH}{d\alpha} \rangle =\frac{1}{\beta}  \bigg(\frac{\partial }{\partial \beta} -  \frac{1}{\beta} \bigg) \frac{\partial Z}{\partial \alpha}, \ \ & \ \  Z\langle JH \rangle =-\sqrt{\ell} \frac{\partial^2 Z}{\partial \beta \partial \mu},\ \ \ \ Z\langle H^2 \rangle = \frac{\partial Z^2}{\partial \beta^2}.
\end{split}
\end{equation}
Substituting \eqref{exphj} in \eqref{part2} yields
\begin{equation}\label{flowpartjt}
\begin{split}
-\frac{2}{\beta}  \frac{\partial Z}{\partial \alpha} +\frac{2\alpha}{\beta}  \frac{\partial^2 Z}{\partial \mu \partial \alpha} - \frac{\alpha^2}{\beta}  \bigg(\frac{\partial }{\partial \beta} -  \frac{1}{\beta} \bigg) \frac{\partial Z}{\partial \alpha} +2 \frac{\partial^2 Z}{\partial \beta \partial \mu} -\alpha \frac{\partial^2 Z}{\partial \beta^2}=0.
\end{split}
\end{equation}
As in the case of $T\bar{T}$ deformation, this can be realized as the defining equation of an analogue of $J\bar{T}$ deformation in quantum mechanics. Equation \eqref{flowpartjt} solves as 
\begin{equation}\label{kerforjt}
Z(\alpha,\beta,\mu)=\int_{\mathcal{C}_\beta} d\beta' \int_\mathcal{C_\mu} d\mu' \ K(\alpha,\beta,\beta',\mu,\mu') \ Z_0(\beta',\mu'),
\end{equation}
where, as before $Z_0(\beta',\mu') $ is the partition function of the undeformed theory at temperature $1/\beta'$ and chemical potential $\mu'$. The contour $\mathcal{C}_\mu$ runs all along the real $\mu'$ axis from $-\infty$ to $\infty$  and the contour $\mathcal{C}_\beta$ runs all along the positive imaginary $\beta'$ axis from $0$ to $i\infty$.  The kernel $K(\alpha,\beta,\beta',\mu,\mu')$  takes the form
\begin{equation}
K(\alpha,\beta,\beta',\mu,\mu') =\frac{\beta}{2\pi i \beta'^2 \alpha} ~\exp \bigg(-\frac{(\mu-\mu')^2}{4\beta'} - \frac{(\beta-\beta')}{\alpha \beta'}(\mu-\mu') \bigg).
\end{equation}
From \eqref{kerforjt} one can read off the deformed spectrum \eqref{spectjt}.

\subsection{$T\bar{T}+J\bar{T}$ deformation in quantum mechanics}\label{TTpJT}

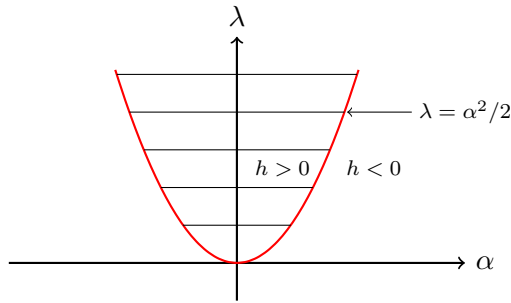
\begin{figure}
\begin{center}
\begin{tikzpicture}[scale=1, transform shape]
\tikzset{middlearrow/.style={
        decoration={markings,
            mark= at position 0.5 with {\arrow{#1}} ,
        },
        postaction={decorate}
    }
}
 \draw[->, thick] (-3,0)--(3,0) node[right]{$\alpha$};
\draw[->, thick] (0,-.5)--(0,3) node[above]{$\lambda$};
\draw[scale=0.5,domain=-3.2:3.2,smooth,variable=\x,red,thick] plot ({\x},{.5*\x*\x});
\draw (-1.414,2)--(1.414,2);
\draw (-1,1)--(1,1);
\draw (-1.58,2.5)--(1.58,2.5);
\draw (-.71,.5)--(.71,.5);
\draw (-1.22,1.5)--(1.22,1.5);
 \draw (.6,1.25) node {\scriptsize{$h>0$}};
 \draw (1.8,1.25) node {\scriptsize{$h<0$}};
 \draw[->] (2.3,2)--(1.45,2) ;
 \draw (3,2) node {\scriptsize{$\lambda=\alpha^2/2$}};
 \end{tikzpicture}
 \caption{The region on and above the red curve in the parameter space corresponds to deformed theories with unitary spectrum while its complement corresponds to those with non-unitary spectrum.}
 \label{figure1}
 \end{center}
\end{figure}

In this subsection, we are going to turn on a deformation that is a general linear combination of $T\bar{T}$ and $J\bar{T}$. As stated in subsection \ref{JTbarqm}, the $J\bar{T}$ deformation alone makes the theory non-unitary. In the discussion that follows, we are going to argue that in the presence of both these deformations, there exist a region in the parameter space (\ie\ the space spanned by $\lambda$ and $\alpha$) where the deformed theory is unitary.  To define such a deformation, we are, as before, going to draw motivation from the deformed spectrum \eqref{TTpJTspect}, \eqref{ABC}. Thus let us define the spectrum of a one-dimensional quantum mechanical system 
with a global $U(1)$ charge $Q$ to be of the form\footnote{Note that, one can further generalize the deformed spectrum by keeping both $\alpha_{\pm}$ and $Q_{L/R}$ from \eqref{TTpJTspect} and defining resulting spectrum as the spectrum obtained by deforming a quantum mechanical system by a general linear combination of three operators namely $T\bar{T}$, $J\bar{T}$ and $T\bar{J}$. The charges $Q_{L/R}$ can be thought of as charges coming from two global $U(1)$'s. }
\begin{equation}\label{ttjtspect}
E(\alpha, \lambda) =-2h\left(1-\frac{\alpha Q}{\sqrt{\ell}} -\sqrt{\left(1-\frac{\alpha Q}{\sqrt{\ell}} \right)^2 + \frac{E_{0}}{h}}~\right),
\end{equation}
where 
\begin{eqnarray}
h=\frac{1}{2\lambda-\alpha^2}
\end{eqnarray}
and as before $\lambda$ and $\alpha$ are respectively the $T\bar{T}$ and $J\bar{T}$ coupling. It's obvious from the deformed spectrum that the theory is unitary in the regime in the parameter space where $h>0$. This corresponds to the region above the red curve (\ie\ the locus $\lambda=\alpha^2/2$) in figure \ref{figure1}. For $h<0$, which corresponds to the region below the red curve  in figure \ref{figure1}, the deformed spectrum above some threshold becomes complex. Approaching the red curve from the above (\ie\ $h\to+\infty$) gives an interesting intermediate regime where the theory is unitary and exhibits many exotic behaviors, that we will discuss later.

Note that in the limit $\alpha\to0$,  \eqref{ttjtspect} gives rise to the $T\bar{T}$ deformed spectrum \eqref{sect1tt} and in the limit $\lambda\to 0$, it gives the $J\bar{T}$ deformed spectrum \eqref{spectjt}. As specified earlier there is an intermediate regime $h\to\infty$ where the deformed spectrum takes the form
\begin{equation}\label{intrspect}
E=\frac{E_0}{1-\frac{\alpha Q}{\sqrt{\ell}}}~.
\end{equation}
The flow equations that would give rise to such a deformed spectrum are given by
\begin{equation}
\begin{split}\label{ttbjtbfloweq}
&\frac{\partial S_E}{\partial \lambda}=-\int d\tau~\frac{T^2}{2\left(1-\frac{\alpha J}{\sqrt{\ell}}\right)+\frac{T}{h}},\\
&\frac{\partial S_E}{\partial \alpha} = \int d\tau~\frac{\alpha T^2+\frac{2JT}{\sqrt{\ell}}}{2\left(1-\frac{\alpha J}{\sqrt{\ell}}\right)+\frac{T}{h}}.
\end{split}
\end{equation}
This leads to the following differential equations in the deformed spectrum:
\begin{equation}\label{flowttjtspec}
\begin{split}
& 2\left(1-\frac{Q\alpha}{\sqrt{\ell}}\right)\frac{\partial E}{\partial \lambda} +\frac{E}{h} \frac{\partial E}{\partial \lambda} +E^2=0,\\
&2\left(1-\frac{Q\alpha}{\sqrt{\ell}}\right)\frac{\partial E}{\partial \alpha} +\frac{E}{h} \frac{\partial E}{\partial \alpha} -\alpha E^2 - \frac{2Q}{\sqrt{\ell}}E=0,
\end{split}
\end{equation}
which solves as \eqref{ttjtspect}.

\subsubsection{Thermal partition function and its flow equation}

Using the same logic as discussed in the case of $T\bar{T}$ and $J\bar{T}$ deformations, the flow equations \eqref{flowttjtspec} for the deformed spectrum imply
\begin{equation}
\begin{split}
& \tr\left[\left(\left(1-\frac{J\alpha}{\sqrt{\ell}}\right)\frac{\partial H}{\partial \lambda} +\frac{H}{h} \frac{\partial H}{\partial \lambda} +H^2\right)e^{-\beta H+\mu J/\sqrt{\ell}}\right]=0,\\
& \tr\left[\left(2\left(1-\frac{J\alpha}{\sqrt{\ell}}\right)\frac{\partial H}{\partial \alpha} +\frac{H}{h} \frac{\partial H}{\partial \alpha} -\alpha H^2 - \frac{2J}{\sqrt{\ell}}H\right)e^{-\beta H+\mu J/\sqrt{\ell}}\right]=0.
\end{split}
\end{equation}
This leads to the following flow equations of the thermal partition function:
\begin{equation}\label{flowpartttjt}
\begin{split}
&2\left(\alpha \frac{\partial }{\partial \mu} -1\right) \frac{\partial Z}{\partial \lambda} +\frac{1}{h}\left(\frac{\partial }{\partial \beta} -\frac{1}{\beta}\right) \frac{\partial Z}{\partial \lambda} +\beta \frac{\partial^2 Z}{\partial \beta^2} = 0,\\
& 2\left(\alpha \frac{\partial }{\partial \mu} -1\right) \frac{\partial Z}{\partial \alpha} +\frac{1}{h}\left(\frac{\partial }{\partial \beta} -\frac{1}{\beta}\right) \frac{\partial Z}{\partial \alpha} -\beta \left( \alpha \frac{\partial}{\partial \beta} - 2 \frac{\partial}{\partial \mu} \right) \frac{\partial Z}{\partial \beta}= 0.
\end{split}
\end{equation}
The above set of differential equations solve as
\begin{equation}\label{kerttjt}
Z(\lambda,\alpha,\beta,\mu)=\int_{\mathcal{C}_\beta} d\beta' \int_\mathcal{C_\mu} d\mu' \ K(\lambda,\alpha,\beta,\beta',\mu,\mu') \ Z_0(\beta',\mu'),
\end{equation}
where, as before $Z_0(\beta',\mu') $ is the partition function of the undeformed theory at temperature $1/\beta'$ and chemical potential $\mu'$. The contour $\mathcal{C}_\mu$ runs along the full imaginary $\mu'$ axis from $-i\infty$ to $i\infty$ and the contour $\mathcal{C}_\beta$ runs along the real positive $\beta'$ axis  from $0$ to $\infty$.
The kernel $K(\lambda,\alpha,\beta,\beta',\mu,\mu')$  takes the form
\begin{equation}\label{ttjtkern}
K(\lambda,\alpha,\beta,\beta',\mu,\mu') =\frac{\beta}{2\pi i \beta'^2 \alpha} ~\exp \bigg(\frac{(\mu'-\mu)^2}{4h\alpha^2\beta'} - \frac{(\beta'-\beta)}{\alpha \beta'}(\mu'-\mu) \bigg).
\end{equation}
For the sake of convergence of the integrals in \eqref{kerttjt}, $h$ has to be greater than zero. For $h<0$, one can define the integral for $h>0$ and then analytically continue the result for $h<0$. Alternatively, for $h<0$, one can choose  appropriate contours of integrations to make the integrals finite yielding the desired deformed partition function. From \eqref{kerttjt}, one can easily read off the deformed spectrum \eqref{ttjtspect}.

\section{Correlation functions and thermodynamics} \label{sec4}

In this section, as an application of the kernel formula \eqref{kerttjt} and \eqref{ttjtkern} discussed in the previous section,  we are going to compute the correlation functions of the deformed theory and analyze the thermodynamics of the deformed system. The discussion that follows, are going to be made in the most general setup namely in the presence of both $T\bar{T}$ and $J\bar{T}$ deformation at the same time. Then we are going to comment on various interesting limiting cases.

\subsection{Deformed correlation function} 

The discussion here is in the same spirit as appears in \cite{Gross:2019uxi}. Let us consider a set of $n$ operators $\mathcal{O}_i(\tau_i)$ $\forall i=\{1,2,\cdots,n\}$ that doesn't change upon deformation inserted at $\tau_i$. For simplicity lets also assume that $\tau_1>\tau_{2}>\cdots>\tau_{n-1}>0$.Then the  n-point thermal correlator of the deformed theory is given by
\begin{equation}
\begin{split}
G^{(n)}(\beta,\{\tau_i\})&=\int dE_1 dQ_1 \langle E_1Q_1|\mathcal{O}_1(\tau_1)\cdots\mathcal{O}_n(0)|Q_1E_1\rangle ~ e^{-\beta E_1+\mu Q_1/\sqrt{\ell}}\\
&= \int\prod_{i=1}^n dE_i dQ_i \langle E_1Q_1|\mathcal{O}_1|Q_2 E_2\rangle \cdots  \langle E_n Q_n|\mathcal{O}_n|Q_1 E_1\rangle ~e^{-\sum_{i=0}^{n-1}\beta_i E_{i+1}+\frac{\mu Q_1}{\sqrt{\ell}}},
\end{split}
\end{equation}
where $\beta_i=\tau_i-\tau_{i+1}$ for $i=0,1,\cdots, n-1$, $\tau_0=\beta$ and $\beta_{n-1}=\tau_{n-1}$. 
Using the kernel formula \eqref{kerttjt} and \eqref{ttjtkern}, one can express the deformed n-point thermal correlation function in terms of the undeformed correlation function $G_0^{(n)}(\beta',\{\tau'_i\})$ as 
\begin{equation}\label{defcorr}
G^{(n)}(\beta,\{\tau_i\})=\int \left(\prod_{i=0}^{n-1} d\beta'_i d\mu'_i ~K(\beta,\beta'_i,\delta_{i,0}\mu,\mu'_i)\right)G_0^{(n)}(\beta',\{\tau'_i\}),
\end{equation}
where $\beta'_i=\tau'_i-\tau'_{i+1}$ for $i=0,1,\cdots, n-1$, $\tau'_0=\beta'$ and $\beta'_{n-1}=\tau'_{n-1}$. 

As a simple example let us consider the deformation of the 2-point functions of conformal quantum mechanics at zero chemical potential. The 2-point functions of operators of dimension $\Delta$ in conformal quantum mechanics is given by \cite{DeAlfaro:1977kq,Chamon:2011xk}
\begin{equation}
\langle\mathcal{O}(\tau)\mathcal{O}(0)\rangle_0=\frac{1}{|\tau|^{2\Delta}}.
\end{equation}
Thus, using \eqref{defcorr} the 2-point function of a conformal quantum mechanics deformed by a general linear combination of $T\bar{T}$ and $J\bar{T}$ at zero chemical potential (\ie\ $\mu=0$) is given by \footnote{In \eqref{def2pt}, $K_n(x)$ is the modified Bessel function of second kind, not to be confused with the kernel $K$.}
\begin{equation}\label{def2pt}
\langle\mathcal{O}(\tau)\mathcal{O}(0)\rangle_{\lambda,\alpha}=2\sqrt{\frac{h}{\pi}}~|\tau|^{-2\Delta+\frac{1}{2}}~e^{2h|\tau|}~K_{2\Delta+\frac{1}{2}}(2h|\tau|).
\end{equation}
Note that the $\lambda$ and $\alpha$ dependences in \eqref{def2pt} come in the combination $h$. Thus, in the limit $h=\infty$ (\ie\ $\lambda=\alpha^2/2$), the two point function of the deformed theory takes the same form as that  of the undeformed theory. This implies that in the absence of chemical potential, there exists a regime in the parameter space (\ie\ for all points on the red curve in figure \ref{figure1}) where the theory ``looks undeformed''. This is reminiscent of the fact that the deformed energy spectrum in the limit $h\to\infty$, at some fixed charge remains unchanged up to some multiplicative factor. 

Starting from a conformal quantum mechanics, the $T\bar{T}$ and the $J\bar{T}$ deformation break certain symmetries (\eg\ scale invarience) of the seed theory. It's likely that in the regime in the parameter space where $h=\infty$, there is a restoration of some amount of broken symmetries. This is reflected in the fact that the 2-point function ``looks undeformed'' in this limit.  It's  as if that the effects of the $T\bar{T}$ and $J\bar{T}$ deformations cancel each other in this regime in the parameter space. In the grand canonical ensemble, where the system is coupled to a finite chemical potential, there we do expect to see some changes. 

For $h<0$, the deformed Euclidean 2-point function \eqref{def2pt} becomes complex. This is related to the non-unitarity of the theory for $h<0$. 

\subsection{Thermodynamics and Hagedorn behavior}
In this subsection we are going to present another application of the kernel formula  \eqref{kerttjt} and \eqref{ttjtkern} in analyzing the thermodynamics of a certain class of quantum systems. This is going to reveal the non-local nature of the theory in the UV.

Let us consider systems with linear specific heat in the presence of a finite chemical potential \footnote{In $d=1$ quantum mechanics, systems like Schwarzian models exhibit linear specific heat \cite{Stanford:2017thb}. The kind of systems we are interested in \eqref{undeformpf} could be thought of as a charged versions of the usual Schwarzian models. This could also be relevant in understanding the holographic dual of $T\bar{T}+J\bar{T}$ deformed charged SYK model.}. The partition function of  such a thermodynamic systems schematically takes the following form
\begin{equation}\label{undeformpf}
\log Z_0 = \frac{c_1}{\beta}+\frac{c_2\mu^2}{\beta},
\end{equation}
where $c_{1,2}$ are system dependent positive constants proportional to the number of degrees of freedom of the system. Using the kernel formula, one can derive the deformed partition function and analyze the thermodynamics of the deformed system. Let's do it in two steps: first lets analyze the $T\bar{T}$ case and then analyze the more general case namely $T\bar{T}+J\bar{T}$ at zero chemical potential of the deformed theory.

\subsubsection{$T\bar{T}$ case}

The  leading contribution to the deformed partition function is given by
\begin{eqnarray}
 \log Z(\lambda,\beta,\beta_h)\sim {\frac{2 c_1}{\beta_h^2}\left(\beta-\sqrt{\beta^2-\beta_h^2}\right)},
\end{eqnarray}
where 
\begin{equation}
\beta_h=\sqrt{2c_1\lambda}
\end{equation}
is the branch point in $\beta$ interpretable as the inverse Hagedorn temperature of the system. Let $F=-\frac{1}{\beta}\log Z(\lambda,\beta,\beta_h)$ be the free energy of the system, then the thermal entropy is given by
\begin{equation}\label{ent}
S=\beta^2 \frac{\partial F}{\partial \beta}\sim \frac{2c_1}{\sqrt{\beta^2-\beta_h^2}}.
\end{equation}
Next, we  Legendre transform $F$ to express the energy $E$ of the system as
\begin{equation}\label{engleg}
E=F+S/\beta.
\end{equation}
From \eqref{ent} and \eqref{engleg} $\beta$ solves as
\begin{equation}\label{betaE}
\beta\sim\frac{(2c_1+\beta_h^2 E)}{\sqrt{4c_1E+\beta_h^2E^2}}.
\end{equation}
Substituting \eqref{betaE} in \eqref{ent}, one finds the thermal entropy of the system as
\begin{eqnarray}
S\sim \sqrt{4c_1E+\beta_h^2E^2}.
\end{eqnarray}
Thus at low energies, the entropy of the system goes as $S\sim \sqrt{4c_1E}$ whereas at very high energies, the entropy grows as $S\sim \beta_h E$ signaling the Hagedorn nature of  the short distance physics with inverse Hagedorn temperature $\beta_h$. 

One can also read of the Hagedorn temperature from the deformed density of states $\rho(\lambda,E)$\begin{equation}
\rho(\lambda,E)\sim e^{\sqrt{4c_1E+\beta_h^2E^2}}.
\end{equation}

\subsubsection{$T\bar{T}+J\bar{T}$ case}\label{sssec4}

The  leading contribution to the deformed partition function, in this case, is given by
\begin{eqnarray}\label{dpartfun}
\log Z(\lambda,\alpha,\beta,\beta_h)\sim  \frac{2 c_1}{\beta_h^2}\left(\beta-\sqrt{\beta^2-\beta_h^{2}}\right).
\end{eqnarray}
Surprisingly the functional form of the deformed partition function at leading order \eqref{dpartfun} is identical to that of the $T\bar{T}$ case.
The inverse Hagedorn temperature, as read off from the branch point of the deformed partition function is given by
\begin{eqnarray}\label{hagt}
\beta_h&=& \sqrt{\frac{c_1}{h}(1+4c_2h\alpha^2)}.
\end{eqnarray}
The inverse temperature and thermal entropy as a function of the energy takes the following form
\begin{equation}\label{entdef}
\begin{split}
&\beta\sim \frac{(2c_1+\beta_h^2 E)}{\sqrt{4c_1E+\beta_h^2E^2}},\\
&S\sim \sqrt{4 c_1E+\beta_h^2E^2}.
\end{split}
\end{equation}
Note that at very high energies, $\beta\to\beta_h$ and $S\sim \beta_h E$, whereas at low energies $S\sim \sqrt{4c_1E}$. The density of states goes as
\begin{eqnarray}\label{rhohag}
\rho(\lambda,\alpha,E)\sim e^{\sqrt{4c_1E+\beta_h^2E^2}}.
\end{eqnarray}

At this point one might wonder which states are contributing to the Hagedorn \eqref{hagt}. To answer this question let us analyze the deformed spectrum \eqref{ttjtspect}. The undeformed energy as a function of the deformed energy and the couplings is given by
\begin{eqnarray}\label{engd}
E_0=\left(1-\frac{\alpha Q}{\sqrt{\ell}}\right)E+\frac{E^2}{4h}.
\end{eqnarray}
The entropy (at fixed charge) of the undeformed theory with partition function \eqref{undeformpf} is given by (see appendix \ref{appA})
\begin{eqnarray}\label{entfq}
S_Q\sim \sqrt{4c_1E_0 -\frac{c_1Q^2}{c_2\ell} }.
\end{eqnarray}
As one varies $\lambda$ and $\alpha$, the energy of the states changes according to \eqref{ttjtspect}, but the number of states remains unchanged. Thus the entropy remains independent of the couplings $\lambda$ and $\alpha$ upon deformation.
Substituting  \eqref{engd} in \eqref{entfq}  one obtains
\begin{eqnarray}\label{ssss}
S_Q\sim \sqrt{4c_1E_0 -\frac{c_1Q^2}{c_2\ell} }=\sqrt{4c_1\left(1-\frac{\alpha Q}{\sqrt{\ell}}\right)E+\frac{c_1E^2}{h}-\frac{c_1Q^2}{c_2\ell}}.
\end{eqnarray}
An interesting fact is that the inverse Hagedorn temperature at fixed charge 
\begin{eqnarray}
\beta^{Q}_h=\sqrt{\frac{c_1}{h}},
\end{eqnarray}
is different from the inverse Hagedorn temperature \eqref{hagt}. This is reminiscent of the fact that the grand canonical ensemble exhibits different Hagedorn behavior compared to fixed charge ensemble. Note that the fixed charge Hagedorn temperature goes to infinity in the limit $h\to\infty$. The thermodynamics (at fixed charge) in this regime is somewhat intermediate between a local theory (\ie\ theories with entropy that goes like $\sqrt{E}$) and a theory with a Hagedorn density of states.
  
At energy $E$, the leading contribution to the fixed charge entropy \eqref{ssss} comes from states with charge
\begin{eqnarray}\label{qes}
Q=-2c_2\alpha\sqrt{\ell}E.
\end{eqnarray}
for which the entropy takes the form
\begin{eqnarray}
S_Q\sim \sqrt{4 c_1E+\frac{c_1}{h}(1+4c_2h\alpha^2)E^2}.
\end{eqnarray}
This is in precise agreement with the second equation in \eqref{entdef} with the inverse Hagedorn temperature given by \eqref{hagt}.

The particular thermodynamic systems that we are interested in, namely systems with linear specific heat  \eqref{undeformpf} exhibit Hagedorn density of states at high energies. It is however important to stress that systems, which do not have linear specific heat (\ie\ are not described by \eqref{undeformpf}), are not likely to exhibit Hagedorn growth. The density of states given in \eqref{rhohag} is {\it{not}} a universal property of $T\bar{T}$ and $J\bar{T}$ deformations in $d=1$; rather this is typical to systems with linear specific heat. 
This is an important difference between two-dimensional and one-dimensional $T\bar{T}+J\bar{T}$ deformation. In $d=2$, the partition function of a generic CFT$_2$, in the thermodynamic limit, exhibits  linear specific heat. Thus, the Hagedorn growth of the density of states of a  $T\bar{T}$ deformed CFT$_2$ at high energies is a universal property unlike $d=1$.

   \subsection{Comments on thermodynamic stability}
   
   In this subsection, we are going to discuss about the thermodynamic stability of a quantum mechanical system with linear specific heat, upon deformation by a general linear combination of $T\bar{T}$ and $J\bar{T}$. We will closely follow the argument in \cite{Barbon:2020amo}.  \footnote{We thank E. Rabinovici for raising the issue of thermodynamic stability which eventually lead to this subsection.}
   
   The temperature of a thermodynamic system at fixed charge is given by
   \begin{eqnarray}\label{cantemp}
   \frac{1}{T(E)}=\frac{\partial S_Q}{\partial E}.
   \end{eqnarray}
   From the monotonicity property of $T(E)$, one can determine the thermodynamic stability of such a thermodynamic system. The sign of $\frac{\partial T}{\partial E}$ determines the sign of the specific heat and systems with positive specific heat (\ie\ $T$ is a monotonically increasing function of $E$) are thermodynamically stable. Systems with negative specific heat (\ie\ $T$ is a monotonically decreasing function of $E$) however, are thermodynamically unstable. In the discussion that follows, we will use this concept to comment of the thermodynamic stability of the deformed system we are interested in. 
   
   We argued in subsubsection \ref{sssec4} that (see eq. \eqref{ssss})
   \begin{eqnarray}\label{sqs0}
   S_Q(E_0(E))=S_Q(E_0).
   \end{eqnarray}
   Differentiating \eqref{sqs0}, with respect to the deformed energy $E$, one can write
   \begin{eqnarray}\label{te}
   T(E)=E'(E_0)T_0(E_0),
   \end{eqnarray}
   where $E'(E_0)$ is the derivative of the deformed energy with respect to the undeformed one and $T_0(E_0)=(\partial S_Q/\partial E_0)^{-1}$ is the temperature of the undeformed theory. For theories with linear specific heat \eqref{undeformpf} \footnote{For thermodynamic stability of systems with non-linear specific heat in the case of $T\bar{T}$ deformation see \cite{Barbon:2020amo}. It would be interesting to study stability of such systems in the case of $T\bar{T}+J\bar{T}$ deformation.}, the undeformed temperature takes the following form
   \begin{eqnarray}\label{t0}
   T_0(E_0)=\frac{1}{2c_1}\sqrt{4c_1E_0 -\frac{c_1Q^2}{c_2\ell} }.
   \end{eqnarray}
   Note that $\frac{dT_0}{dE_0}$ is always positive implying that the seed theory is thermodynamically stable.
   Substituting \eqref{t0} in \eqref{te} and using the deformed spectrum \eqref{ttjtspect}, one can write
   \begin{eqnarray}\label{dtde}
   \frac{\partial T}{\partial E}=\frac{Q^2+4c_2h\ell\left(1-\alpha\frac{Q}{\sqrt{\ell}}\right)^2}{4c_2\ell \sqrt{4c_1E_0 -\frac{c_1Q^2}{c_2\ell} }\left(E_0+h\left(1-\alpha\frac{Q}{\sqrt{\ell}}\right)^2\right)}.
   \end{eqnarray}
   It is easy to see  that for $h>0$, (\ie\ for the regime in the parameter space where the deformed theory is unitary),  $\frac{\partial T}{\partial E}>0$ (remember that $c_2>0$). Thus the deformed theory has a positive specific heat implying that the theory is  thermodynamically stable.
   
   Next, let us discuss what happens for $h<0$ where the theory is non-unitary. Assuming reality of \eqref{entfq} (\ie\ assuming $E_0>\frac{Q^2}{4c_2\ell}$), the deformed system has positive specific heat for 
   \begin{eqnarray}\label{hcond}
   0>h>-\frac{Q^2}{4c_2\ell \left(1-\alpha\frac{Q}{\sqrt{\ell}}\right)^2}\ \ \ \text{ OR }\ \ \ h<-\frac{E_0}{\left(1-\alpha\frac{Q}{\sqrt{\ell}}\right)^2},
   \end{eqnarray}
   for a specified $E_0$ and $Q$.
   Thus, for $h<0$, the system is thermodynamically stable provided $h$ satisfies \eqref{hcond}. Note that, for non-unitary theories, the stability is highly state dependent.

   \section{UV completion}\label{sec5}
   
   It was shown in \cite{Gross:2019ach} that the theory defined by the flow equation \eqref{ttbarflow} has a non-perturbative UV completion: quantum mechanics coupled to  gravity in $d=1$. The idea of such UV completion has its origin in \cite{Dubovsky:2017cnj,Dubovsky:2018bmo} in the case of $T\bar{T}$ deformation in $d=2$ and was further generalized in \cite{Aguilera-Damia:2019tpe} for a deformation by a general linear combination of $T\bar{T}, J\bar{T}$ and $T\bar{J}$  also in $d=2$. In the discussion that follows, we will briefly describe the non-perturbative UV completion of $T\bar{T}$ deformation in $d=1$ discussed in  \cite{Gross:2019ach} and analogously propose a non-perturbative UV completion  for a more general deformation of a quantum mechanical system defined by the flow equations \eqref{ttbjtbfloweq}.
   
   \subsection{$T\bar{T}$ case}
   
   The UV completion proposed in \cite{Gross:2019ach} is the following:
   \begin{eqnarray}
   Z(\lambda,\beta)=\int \frac{De DX D\Phi}{V_{\rm{diff}}}~ \exp\left(-S_0[e,\Phi]-S[\lambda;e,X]\right)~,
   \end{eqnarray}
   where $S_0[e,\Phi]$ is the action of the seed theory which contains some matter fields, collectively denoted by $\Phi$, coupled to one-dimensional gravity with einbein $e(\tau)$. The Euclidean time is compact $\tau\sim\tau+\beta'$ with periodicity $\beta'$. The full deformed theory is invariant under reparametrization of $\tau$. Thus, in the path integral, one needs to divide by the volume of the group of diffeomorphisms denoted by $V_{\rm{diff}}$. The reparametrization invariant $d=1$ gravity action is given by
   \begin{eqnarray}\label{1dgrav}
   S[\lambda;e,X]=\frac{1}{2\lambda}\int_0^{\beta'} d\tau e\left(\frac{\dot{X}}{e}-1\right)^2,
   \end{eqnarray}
   where $X$ (with $\dot{X}\equiv \partial_\tau X$) is a compact bosonic field with periodicity $\beta$ that satisfies $X(\tau+\beta')=X(\tau)+m\beta$ with $m\in \mathbb{Z}$ is the winding of $X$ along the $\tau$ circle.  On fixing the einbein gauge by introducing Faddeev-Popov ghosts and choosing the gauge $e=1$ and finally dividing by the volume of the residual symmetry, the deformed partition function takes the following form:
   \begin{eqnarray}
   Z(\lambda,\beta)= \frac{\beta}{\sqrt{2\pi\lambda}} \int_{0}^\infty \frac{d\beta'}{\beta'^{3/2}} \sum_{m\in\mathbb{Z}}\exp\bigg(-\frac{(m\beta-\beta')^2}{2\beta'\lambda} \bigg)~ Z_0(\beta'),\label{defgpar}
   \end{eqnarray}
   where the undeformed partition function $Z_0(\beta')$ is given by
   \begin{eqnarray}
   Z_0(\beta')=\int D\Phi ~ \exp\left(-S_0[e=1,\Phi]\right).
   \end{eqnarray}
   The partition function \eqref{defgpar} of the deformed theory in the unit winding sector matches exactly with the deformed partition function \eqref{kerfortt} and \eqref{kertt} obtained from the kernel analysis.
   
 \subsection{$T\bar{T}+J\bar{T}$ case}
 
 Following the discussion in the previous section, one may guess that the non-perturbative definition of a $T\bar{T}+J\bar{T}$ deformed quantum mechanical system governed by the flow equation \eqref{ttbjtbfloweq} could be obtained by coupling the theory to gravity as well as a $U(1)$ gauge field in $d=1$. For deatailed discussion on gauge and gravity theory in $d=1$ see \cite{Elitzur:1992bf}.  In the following discussion we will perform the path integral over the additional fields (\ie\ the einbein, the gauge field and two compact real scalar fields) that will lead to the Kernel formula \eqref{kerttjt} and \eqref{ttjtkern}
 
We propose 
 \begin{eqnarray}\label{pig}
 Z(\lambda,\alpha,\beta)=\int \frac{De DX Df DY D\Phi}{V_{\rm{diff}}\times V_{\rm{gauge}}} ~\exp\left(-S_0[e,f,\Phi]-S[\lambda,\alpha;e,X,f,Y]\right) ~,
 \end{eqnarray}
 as the non-perturbative UV completion of the deformed theory in the most general case.
 The fields $e$ and $X$ are, as before, the einbein and the compact scalar field with periodicity $\beta$, $f$ is the $U(1)$ gauge field and $Y$ is a compact scalar with periodicity $\mu'$ satisfying $Y(\tau+\beta')=Y(\tau)+\tilde{m}\mu'$ with $\tilde{m} \in \mathbb{Z}$. The action of the seed theory coupled to gravity and gauge field is denoted  by $S_0[e,f,\Phi]$. This coupling is done in a way such that $S_0[e,f,\Phi]$  respects invariance under  reparametrization in $\tau$ and $U(1)$ gauge symmetry.  In fact, the full deformed theory should be invariant under reparametrization and gauge symmetry as a result of which, one needs to divide by the volume of the diffeomorphism (denoted by $V_{\rm{diff}}$) and gauge group (denoted by $V_{\rm{gauge}}$). The reparametrization and gauge invariant $d=1$ gravity plus gauge theory action is given by \footnote{Note that, pure gauge and gravity theory in $d=1$ have negative degrees of freedom. To make the theories physically meaningful one needs to add the scalar fields $X$ and $Y$ resulting in a theory with $0$ degrees of freedom \cite{Elitzur:1992bf}. Thus one may view the theory solely given by \eqref{sact} as topological.}
\begin{eqnarray}\label{sact}
   S[e,X,f,Y]=\int_0^{\beta'} \frac{d\tau}{e}\left[-\frac{1}{4h\alpha^2}(\dot{Y}-f)^2-\frac{1}{\alpha}(\dot{Y}-f)(\dot{X}-e)\right].
   \end{eqnarray}   
  Under reparametrization $\tau\to\tilde{\tau}$, the fields $(e,\dot{X},f,\dot{Y})$ transform as
  \begin{eqnarray}
  \begin{split}\label{rept}
 & e(\tau)\to\tilde{e}(\tilde{\tau})=\left(\frac{\partial\tau}{\partial\tilde{\tau}}\right)e(\tau),\\
 & \partial_\tau X(\tau)\to\partial_{\tilde{\tau}}\tilde{X}(\tilde{\tau})= \left(\frac{\partial\tau}{\partial \tilde{\tau}}\right) \partial_{\tau} X(\tau),\\
 & f(\tau)\to\tilde{f}(\tilde{\tau})=\left(\frac{\partial\tau}{\partial\tilde{\tau}}\right)f(\tau),\\
 & \partial_\tau Y(\tau)\to\partial_{\tilde{\tau}}\tilde{Y}(\tilde{\tau})= \left(\frac{\partial\tau}{\partial \tilde{\tau}}\right) \partial_{\tau} Y(\tau),\\
   \end{split}
  \end{eqnarray}
   and under gauge transformation the fields $(f,\dot{Y})$ transform as
 \begin{eqnarray}
     \begin{split}\label{gt}
 & f\to f+g, \\
 & \dot{Y} \to \dot{Y} +g,
     \end{split}
   \end{eqnarray}  
   where $g$ is some arbitrary function of $\tau$. 
 One can easily check that the action  \eqref{sact} is invariant under reparametrization symmetry \eqref{rept} and gauge transformation \eqref{gt}.
 
 Next let's fix gauge. We set $e=1$ with $\tau\sim \tau+\beta'$. This fixes almost all the reparametrization symmetry except constant shifts in $\tau$ and $\tau\to-\tau$ symmetry. To fix this residual symmetry we will divide the path integral by the volume of the this residual gauge group which turns out to be $2\beta'$.\footnote{The volume of the residual gauge symmetry is simply twice the circumference of the $\tau$ circle. The 2 factor comes form $\tau\to-\tau$ symmetry.} Similarly let's choose $f=0$ with $\beta' f \sim \beta' f+\mu'$ to fix the $U(1)$ gauge symmetry. As in the case of reparametrization symmetry, this gauge fixing keeps unfixed the symmetry due to constant shifts of the zero mode of $f$ and its $\mathbb{Z}_2$ sign flip. To fix this residual gauge symmetry we divide by $2\mu'$.\footnote{It is easy to convince oneself from dimensional analysis that $2\mu'$ is indeed the appropriate factor to divide the path integral by in order to kill the residual gauge symmetry. For a more accurate treatment one may need to adapt the techniques used in string worldsheet theory to fix residual worldsheet gauge symmetry. This has been illuminated in footnote 3 of \cite{Gross:2019ach}.} 
 
 To fix the einbein gauge and the $U(1)$ gauge symmetry we adopt the Faddeev-Popov procedure.  The Faddeev-Popov measures for the reparametrization and $U(1)$ gauge symmetry is defined as 
 \begin{eqnarray}
 \begin{split}\label{fpd}
& 1=\Delta_{FP}(e)\int_0^\infty d\beta' \int D\zeta ~\delta(e-1^\zeta) ,\\
&  1=\Delta_{FP}(f)\int_{-\infty}^\infty d\mu' \int D\eta ~\delta(f-0^\eta), 
 \end{split}
 \end{eqnarray}
where $\zeta$ is the diffeomorphism in $\tau$ and $1^\zeta$ is a diffeomorphism transformation of the fudicial einbein $e=1$ and similarly $\eta$ is the $U(1)$ gauge transformation and $0^\eta$ is the gauge transformed fudicial gauge field $f=0$. Next, we follow the usual procedure of writing the delta functions in \eqref{fpd} as Fourier integrals and in introducing Grassmann fields $a_i,b_i$ and $c_i$ for $i=1,2$ to invert the Faddeev-Popov measure. The Faddeev-Popov measures can be expressed as \footnote{For the ghost fields, we use the same normalization as in \cite{Gross:2019ach}.}
\begin{eqnarray}
\begin{split}\label{fpg}
& \Delta_{FP}(e)=\int_0^\infty d \beta'  \int Da_1 Db_1 Dc_1 ~\exp\left(-4\int_0^{\beta'} d\tau (b_1\dot{c}_1-a_1b_1/\beta')\right),\\
& \Delta_{FP}(f)=\int_0^\infty d \beta'  \int Da_2 Db_2 Dc_2 ~\exp\left(-4\int_0^{\beta'} d\tau (b_2\dot{c}_2-a_2b_2/\beta')\right).
\end{split}
\end{eqnarray}
Next we substitute \eqref{fpd} and \eqref{fpg} in \eqref{pig} and perform the path integrals over the $a_i$ ghost fields. This gives
\begin{eqnarray}
\begin{split}\label{fullpf}
Z(\lambda,\alpha;\beta)&=\int_0^\infty d\beta' \int_{-\infty}^\infty d\mu' \int \prod_{i=1}^2 [Db_i Dc_i] D\Phi DX DY\prod_{i=1}^2\left(\frac{4}{\beta'}\int_0^{\beta'} d\tau ~b_i\right) \\
&\times \exp\left(-S_0[e=1,f=0,\Phi]-S[\lambda,\alpha;e=1,X,f=0,Y]-4\sum_{i=1}^2\int_0^{\beta'}d\tau ~b_i\dot{c}_i \right).
\end{split}
\end{eqnarray}

   Fourier expansion of $X$ and $Y$ fields are given by
   \begin{eqnarray}
   \begin{split}\label{xym}
 &  X(\tau)=\tau\left(\frac{m\beta}{\beta'}\right)+\frac{1}{\sqrt{\beta'}}\sum_{n=-\infty}^{\infty}\exp\left(2\pi i n\tau/\beta'\right)X_n,\\
 & Y(\tau)=\tau\left(\frac{\tilde{m}\mu'}{\beta'}\right)+\frac{1}{\sqrt{\beta'}}\sum_{n=-\infty}^{\infty}\exp\left(2\pi i n\tau/\beta'\right)Y_n,
      \end{split}
   \end{eqnarray}
   where the reality condition on the fields $X$  and $Y$ demands $X_n=X_{-n}^\ast$ and $Y_n=Y_{-n}^\ast$ 
   Since $X$ and $Y$ are compact bosons, the zero modes $X_0$ and $Y_0$ satisfies the periodicity condition $X_0\sim X_0+\sqrt{\beta'}\beta$ and $Y_0\sim Y_0+\sqrt{\beta'} \mu'$. 
   Next, we substitute the mode expansions \eqref{xym} in the gauge fixed action \eqref{sact}, which gives
   \begin{eqnarray}
   S[e=1,X,f=0,Y]=S_{w}+S_{osc},
   \end{eqnarray}
   where $S_{w}$ and $S_{osc}$ denote respectively the contribution from the winding mode and the oscillators given by
   \begin{eqnarray}
   \begin{split}\label{wosc}
 &  S_{w}=-\frac{\tilde{m}^2\mu'^2}{4h\alpha^2\beta'}-\frac{\mu'}{\alpha\beta'}(m\beta-\beta'),\\
 &  S_{osc}=-\frac{2\pi^2}{4h\alpha^2\beta'^2}\sum_{n=-\infty}^\infty n^2\left(Y_nY_{-n}+4h\alpha Y_nX_{-n}\right).
    \end{split}
   \end{eqnarray}
   The path integral of the oscillator part of the action \eqref{wosc} can be evaluated exactly:
   \begin{eqnarray}\label{infp}
   \int DX DY ~\exp\left(-S_{osc}\right)=\int_{0}^{\beta\sqrt{\beta'}} dX_0\int_{0}^{\mu'\sqrt{\beta'}} d Y_0 \prod_{n=1}^{\infty}\left(-\frac{h\alpha^2\beta'^2}{\pi n^2}\right)\prod_{n=1}^\infty
\left(\frac{\beta'^2}{4\pi h n^2}\right) . 
 \end{eqnarray}
 The integrals over the zero modes namely $X_0$ and $Y_0$ are ordinary definite integrals and can easily be evaluated.
 The infinite products in \eqref{infp} can be explicitly computed using zeta-function regularization techniques (see appendix \ref{appB} for details) giving rise to 
 \begin{eqnarray}
 \int DX DY ~\exp\left(-S_{osc}\right)= \frac{\beta\mu'}{2\pi i \beta'\alpha}.
 \end{eqnarray}
 
The path integral over the non-zero modes of the ghost fields are given by
\begin{eqnarray}
\int Db_i Dc_i ~\exp\left(-4\int_0^{\beta'} d\tau b_i\dot{c}_i\right)=\prod_{n=1}^{\infty}\frac{8\pi n}{\beta'}=\frac{\sqrt{\beta'}}{2}.
\end{eqnarray}
In the ghost path integral, we have excluded the integrals over the zero modes of the $c_i$ ghost fields. The integrals of the zero modes of the $c_i$ ghost fields takes care of redundancies associated with the residual gauge symmetries. Since have excluded the integrals over the zero modes of the $c_i$ ghost fields, one needs to divide out by the volume of the residual gauge groups (both  residual reparametrization symmetry and residual $U(1)$ gauge symmetry).

Next, one needs to compute the integrals over the zero modes of the $b_i$ ghost fields. Since $b_i$ is periodic in $\tau$ with periodicity $\beta'$, the integrated $b_i$ insertions in \eqref{fullpf} picks out contribution only from the zero modes $b_i^0$. Thus the contribution from the integral over the zero mode of $b_i$ ghost fields is given by  
\begin{eqnarray}
\int db_i^0\left(\frac{4}{\beta'}\int_0^{\beta'} d\tau ~b_i\right) =\frac{4}{\sqrt{\beta'}} \int db_i^0 b_i^0=\frac{4}{\sqrt{\beta'}}.
\end{eqnarray}
 
 Thus putting together all the pieces and dividing by the volume of the residual symmetry groups one obtains
 \begin{eqnarray}\label{zzz}
 Z(\lambda,\alpha;\beta)=\frac{\beta}{2\pi i \alpha}\int_{-\infty}^\infty d\mu' \int_0^\infty \frac{d\beta'}{\beta'^2} \sum_{m,\tilde{m}}\exp\left(\frac{\tilde{m}^2\mu'^2}{4h\alpha^2\beta'}+\frac{\tilde{m}\mu'}{\alpha\beta'}(m\beta-\beta')\right)Z_0(\beta',\mu').\nonumber\\
 \end{eqnarray}
 Note that the above integral is strictly convergent for $h<0$. For $h>0$ we choose the $\mu'$-contour to run along the imaginary $\mu'$ axis from $-i\infty$ to $+i\infty$. With this choice of the $\mu'$-contour and for unit windings, $m=\tilde{m}=1$, \eqref{zzz} agrees precisely with the kernel formula \eqref{kerttjt} with \eqref{ttjtkern} at $\mu=0$. For non-zero $\mu$ all one needs to do is to send $\mu'$ to $\mu'-\mu$ in \eqref{zzz}.

\section{Discussion}\label{sec6}

The main goal of this paper is to continue the study of an analogue of $T\bar{T}$ deformation of a $d=1$ quantum mechanical system introduced in \cite{Gross:2019ach} and further generalize such solvable deformations in $d=1$ to those that have properties similar to $J\bar{T}$ deformation and deformation by an arbitrary linear combination of $T\bar{T}$ and $J\bar{T}$ in $d=2$. Unlike the study of $T\bar{T}$ in quantum mechanics, which has its motivation from JT gravity in $AdS_2$ with a Dirichlet wall at a finite radial distance, we draw motivation from the brunch cut singularity of the spectrum in $d=2$ to define solvable deformations in quantum mechanics that can be thought of as an analogue of $J\bar{T}$ and $T\bar{T}+J\bar{T}$ deformations in $d=1$. We have shown that the thermal partition function of such deformed theories satisfy a flow equation whose solution gives the deformed partition function as an integral of the undeformed partition function weighted by some kernel. We argued that this can be realized as an alternative definition of the deformations valid even at quantum level.

The kernel formula has various interesting applications. It has been used to compute the deformed two-point functions. One intriguing observation is that in the presence of $T\bar{T}+J\bar{T}$ deformation, there is a regime (\ie\ $2\lambda=\alpha^2$) in the parameter space  where the two-point function looks undeformed. It's likely that in this regime in the parameter space there is an enhancement of symmetries. We would like to leave the detailed analysis of the symmetries of the system in this limit as a future exercise. The kernel formula can further be used in analyzing the thermodynamics of the deformed system. We concluded that for a deformed quantum mechanical system with undeformed partition function given by \eqref{undeformpf}, the microscopic origin of the states with Hagedorn density \eqref{hagt} arises from the charged states  \eqref{qes}. We also studied the thermodynamic stability of such deformed systems and found that if is the deformed theory is unitary, it is also thermodynamically stable. For the regime in the parameter space where the deformed theory is non-unitary, thermodynamic stability is highly state dependent.

Our study of irrelevant deformations of one-dimensional quantum mechanics pave the way to various direction of future research, since the arena of $T\bar{T}$ and $J\bar{T}$ deformation in $d=1$ is highly unexplored. An obvious extension of our work would be to extend the computation of deformed correlation functions in section \ref{sec4} to higher point functions. We studied two-point function of operators of the deformed theory that do not depend on the couplings explicitly. It would be interesting to extend our analysis to those operators that have explicit dependence on the couplings \eg\ it would be nice to calculate two-point functions of the stress tensor operator and the $U(1)$ current. 

Perhaps the most interesting extension of our work would be to understand the holographic realization of $J\bar{T}$ deformation in quantum mechanics and its relation to charged SYK model. Given that the flow equation of the partition function of a $J\bar{T}$ deformed quantum mechanical system is known, it would be interesting to understand, what kind of two-dimensional bulk gravity yields a similar flow equation. A possible way to proceed would be to understand $J\bar{T}$ deformation of charged SYK model. This may lead us to some deformation of JT gravity (possibly with some matter fields turned on) with a cutoff  and mixed boundary conditions, the flow equation of which resembles the flow equation of $J\bar{T}$ deformed quantum mechanics. Eventually one may be interested in understanding the bulk dual upon turning on $T\bar{T}+J\bar{T}$. 

Other interesting open problems include understanding the entanglement structure of these deformed theories. One may also be interested in computing the deformed Lagrangian of the some simple quantum mechanical systems (\eg\ single real boson with polynomial potential \etc.). It would also be interesting to perform supersymmetric extensions of these deformations.

\section*{Acknowledgements} 

We would like to thank Eliezer Rabinovici for helpful comments on the manuscript.
The work of SC and AM are supported by the Infosys Endowment for the study of the Quantum Structure of Spacetime.

\appendix
\section{Entropy and partition function}\label{appA}

In this appendix we are going to show that  if the fixed charge thermal entropy is given by
\begin{equation}\label{entropyE0Q}
S(E,Q)\sim \sqrt{4c_1E -\frac{c_1Q^2}{c_2\ell} },
\end{equation}
then this must correspond to a theory with grand canonical partition function given by \eqref{undeformpf}.

The grand canonical partition function of a quantum mechanical system with $U(1)$ global symmetry is given by 
\begin{equation}\label{pardef}
Z =\int dE \int dQ ~ \rho(E,Q) ~ \exp\left(-\beta E +\mu Q/\sqrt{\ell} \right) ,
\end{equation}
where $\rho(E,Q)=e^{S(E,Q)}$ is the density of sates at fixed energy $E$ and charge $Q$. We perform the $E$ and $Q$ integrals by saddle point approximation. The saddles are located at
\begin{eqnarray}
\begin{split}\label{saddle}
&E=\frac{c_1+c_2\mu^2}{\beta^2},\\
& Q=\frac{2c_2\sqrt{\ell}\mu}{\beta}.
\end{split}
\end{eqnarray}
Substituting the saddles \eqref{saddle} in \eqref{pardef} one obtains
\begin{equation}
Z \sim \exp\left(\frac{1}{\beta}(c_1+c_2\mu^2)\right). 
\end{equation}

\section{Zeta function regularization}\label{appB}

For a positive non-decreasing infinite sequence $0< \lambda_1 \leq \lambda_2 \leq \dots $, the zeta-regularized product of the infinite sequence is defined as \cite{Mizuno}
\begin{equation}\label{rzp}
\prod_{n=1}^{\infty} \lambda_{n} = \exp\left(-\zeta'_{\lambda}(0) \right),
\end{equation}
where 
\begin{equation}\label{gzeta}
\zeta_{\lambda}(s) = \sum_{n=1}^{\infty} ~ \frac{1}{\lambda_{n}^s}
\end{equation}
is the generalized zeta function and $\zeta'_{\lambda}(s)\equiv d\zeta_{\lambda}(s)/ds$. In the case $\lambda_n=n$ where $n\in\{1,2,\cdots\}$, \eqref{gzeta} reduces to the ordinary Riemann zeta function $\zeta(s)$.

Using the zeta-regularized product formula \eqref{rzp}, we would like to compute the following infinite product:
 \begin{equation}\label{ns}
{N}_{s}=\prod_{n=1}^{\infty}( {n}^s a),
\end{equation}
where $a$ is a c-number. Taking logarithm on either side of \eqref{ns} one can write 
\begin{equation}\label{logns}
\log {N}_{s}=\zeta(0) \log a + s\log \left(\prod_{n=1}^{\infty} {n} \right).
\end{equation}
Next, we use the fact that $\zeta(0)=-\frac{1}{2}$ and $\zeta'(0)=-\frac{1}{2}\log 2\pi$, and use \eqref{rzp} to write
\begin{equation}\label{infpn}
\prod_{n=1}^{\infty} {n} = \sqrt{2\pi}. 
\end{equation}
Substituting \eqref{infpn} in \eqref{logns}, one obtains
\begin{equation}
N_{s} = \sqrt{\frac{(2\pi)^s}{a}}.
\end{equation}

\newpage


\providecommand{\href}[2]{#2}\begingroup\raggedright\endgroup

\end{document}